\documentclass[conference]{IEEEtran}
\IEEEoverridecommandlockouts
\usepackage{cite}
\usepackage{amsmath,amssymb,amsfonts}
\usepackage{algorithmic}
\usepackage{graphicx}
\usepackage{subfigure}
\usepackage{textcomp}

\usepackage{amsmath,amssymb,amsfonts}
\usepackage{xcolor}
\def\BibTeX{{\rm B\kern-.05em{\sc i\kern-.025em b}\kern-.08em
    T\kern-.1667em\lower.7ex\hbox{E}\kern-.125emX}}
\begin{document}

\title{
Enhancing Wearable based Real-Time Glucose Monitoring via Phasic Image Representation Learning based Deep Learning
}

\author{\IEEEauthorblockN{\textsuperscript{1} Yidong Zhu, \textsuperscript{1} Nadia B Aimandi, \textsuperscript{1, 2, 3} Mohammad Arif Ul Alam}
\IEEEauthorblockA{\textsubscript{1}\textit{Computer Science, University of Massachusetts Lowell}}
\IEEEauthorblockA{\textsubscript{2}\textit{Medicine, University of Massachusetts Chan Medical School}}
\IEEEauthorblockA{\textsubscript{3}\textit{National Institute on Aging, National Institute of Health}}
}

\maketitle

\begin{abstract}
In the U.S., over a third of adults are pre-diabetic, with 80\% unaware of their status. This underlines the need for better glucose monitoring to prevent type 2 diabetes and related heart diseases. Existing wearable glucose monitors are limited by the lack of models trained on small datasets, as collecting extensive glucose data is often costly and impractical. Our study introduces a novel machine learning method using modified recurrence plots in the frequency domain to improve glucose level prediction accuracy from wearable device data, even with limited datasets. This technique combines advanced signal processing with machine learning to extract more meaningful features. We tested our method against existing models using historical data, showing that our approach surpasses the current 87\% accuracy benchmark in predicting real-time interstitial glucose levels.

\end{abstract}


\section{Introduction}
In the United States, approximately 34.5\% of adults aged 18 years or older, equating to 88 million individuals, are estimated to have pre-diabetes, with a staggering 90\% of them being unaware of their condition \cite{cdc2020}. This state of unawareness is particularly concerning given that pre-diabetes, a precursor to more severe health conditions like type 2 diabetes, can often be mitigated or reversed through early lifestyle interventions such as dietary changes, regular physical activity, and maintaining a healthy weight \cite{ada202}. Despite this potential for reversal, studies indicate that without these interventions, 15-30\% of people with pre-diabetes will progress to type 2 diabetes within five years \cite{ada202}. The annual conversion rate from pre-diabetes to diabetes is alarmingly high, estimated at around 10\%, underscoring the urgent need for effective and continuous monitoring mechanisms \cite{cdc2020}. This transition not only poses significant health risks but also contributes to the increasing healthcare burden associated with diabetes management.


The healthcare industry currently grapples with the significant challenge of the absence of non-invasive, easily accessible methods for glucose monitoring, a key factor in the effective self-management of pre-diabetes. Traditional tools such as blood glucose meters and continuous glucose monitors, while useful, often involve invasive procedures and can be cost-prohibitive, limiting their widespread adoption \cite{zhang19}. In recent years, there has been a notable rise in the popularity of wearable technologies, particularly wrist-worn biometric devices \cite{statista21, patel15, li20}. These devices have seen remarkable market penetration, with over 117 million units currently in use, and projections suggest a potential doubling in the next few years \cite{statista21}. These wearables are evolving to track more than just basic health metrics, positioning themselves as pivotal in the discovery of digital biomarkers, which are crucial in transforming extensive health data into actionable insights \cite{patel15}. The role of such biomarkers is becoming increasingly significant in healthcare, particularly in their potential to significantly reduce the incidence of pre-diabetes through early detection and management \cite{li20}.

In our study, we present a novel approach that involves generating RGB images from wearable signal data, including skin temperature, electrodermal activity, and blood volume pulse, using modified recurrence plots in phasic domain, and then employing deep learning algorithms pre-trained on extensive image datasets to enhance blood glucose level detection.

Our {\bf key contributions:}
\begin{figure}[!htb]
\begin{center}
 \includegraphics[width=\linewidth]{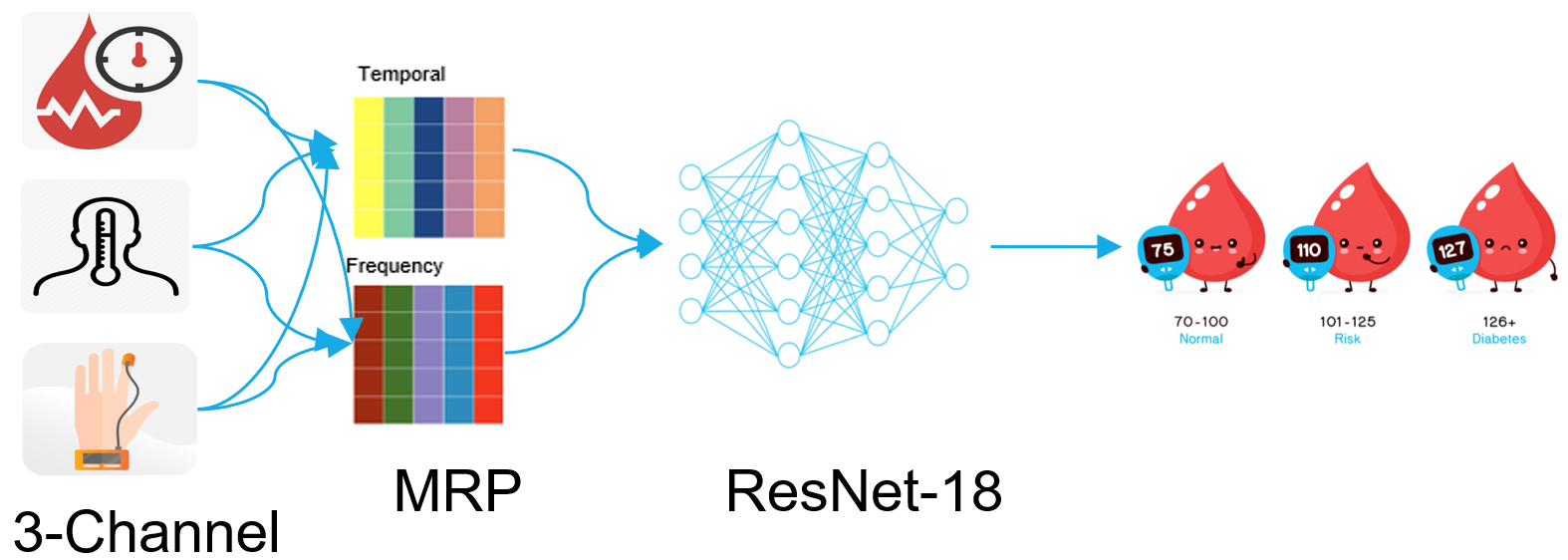}
 \caption{The schematic diagram of proposed work}
 \label{fig:MRP}
\end{center}
\end{figure}
\begin{itemize}
\item Our research introduces the Phasic Recurrent Plot (PRP), a new technique for transforming multi-modal wearable signals into Recurrence Plot (RP) Images. This method replaces existing `temporal' RPs that represent time-sequenced data patterns, while proposed 'phasic' RPs emphasize frequency variations in these multi-modal signals.

\item Our study implemented a pre-trained deep learning model for image processing to analyze RP images of heart rate, electrodermal activity, and temperature data collected from 16 individuals over 8-10 days, exceeding existing accuracy benchmarks in real-time interstitial glucose level prediction.

\end{itemize}

\section{Related Works}



\subsection{Recurrent Plot for Wearable Signal Representation}
Previous research in the field of wearable technology has made significant strides in utilizing Recurrence Plots (RPs) for the analysis of nonlinear dynamical systems, as highlighted in the tutorials for behavioral sciences by Webber and Zbilut \cite{webber2005}. Gao and Cai \cite{gao2013} further developed RP techniques by integrating them with machine learning algorithms, enhancing pattern recognition capabilities in wearable data. Li et al. \cite{li2019} applied RPs in real-time analysis of electrocardiogram (ECG) signals, demonstrating their utility in wearable health monitoring. However, these studies primarily focused on temporal domain analysis and single-modal data, as evidenced in the works of Romano et al. \cite{romano2007} and Silva and Hidalgo \cite{silva2015}, respectively. Our proposed method, in contrast, innovatively fuses three-modal data in the phasic domain, overcoming the limitations of previous systems by offering a more comprehensive and accurate analysis for wearable sensor data, marking a significant advancement in wearable technology and health monitoring.

\subsection{Non-Invasive Glucose Monitoring}
The predictive potential of noninvasive wearable technologies for blood glucose levels has been extensively explored through various biological signals and parameters, including photoplethysmography (PPG) signal \cite{reference4}, sleep and physical activity data \cite{reference5}, and techniques like infrared spectroscopy\cite{infrared}, ultrasound \cite{ultrasound}, and fluorescence \cite{fluorescence}. However, these methods face challenges such as lack of specificity, skin irritation, and significant time lags in interstitial fluid glucose measurement, coupled with poor correlation with actual blood glucose levels and instability due to external factors. The success of non-invasive glucose estimation hinges on sophisticated signal processing and noise reduction, requiring extensive datasets for machine learning and deep learning algorithm development. A notable instance is the Duke University Health System's study, which combined food diaries with noninvasive wearables, including an invasive Dexcom 6 CGM and an Empatica E4 wristband, worn by participants aged 35-65. This study generated a dataset of 25,000 concurrent readings from glucose monitors and smartwatches \cite{reference2}, employing traditional feature extraction and machine learning techniques but faced limitations in achieving high accuracy. Utilizing a combination of domain-driven and data- driven feature engineering, the study crafted 69 variables that incorporated multiple factors influencing glucose levels, including diet, stress, physical activity, and circadian rhythm. A decision tree classifier, leveraging this rich feature set, exhibited a balanced accuracy of 84.3\% in detecting personalized glucose deviations \cite{reference1}. In our paper, we introduce the Phasic Recurrent Plot (PRP) technique, which converts multi-modal wearable sensor data into 'phasic' Recurrence Plot (RP) Images, leveraging pre-trained computer vision models to boost the accuracy of wearable-based glucose estimation. This novel approach effectively transforms frequency-variant data from wearable sensors into an image format suitable for advanced image processing.

\section{Frequency Domain Information Encoding in Image}
\subsection{Visualization via Recurrence Plots}
A Recurrence Plot (RP) serves as a graphical tool to analyze complex dynamic systems, effectively capturing phase space trajectories of nonlinear data \cite{reference25}. This visualization technique plots small-scale features as dots and lines and large-scale patterns like homogeneity and drift. Formally, the RP is defined as a matrix $\mathbf{R}$, derived from a set of trajectory data $\mathbf{x}$, where each matrix element $R_{i,j}(\varepsilon)$ is the L2 norm of the difference between the trajectory points $\mathbf{x}_i$ and $\mathbf{x}_j$. This method is adept at transforming 3-axis signal data into the RGB channels of an image. We represent states in the phase space as \(s_j = (x_j, x_{j+1})\), with \(s_j \in \mathbb{R}^2\). The RP is constructed using a matrix \(R\), where \(R \in \mathbb{R}^{(N-1) \times (N-1)}\), and each element reflects the L2 norm of the difference between states. The RP matrix is articulated as:
\begin{equation}
\label{eq:rp_matrix}
    R_{m,n} = ||s_i - s_j||
\end{equation}

\subsection{Temporal Adjustment in Recurrent Plots}
The inherent symmetry of the recurrence matrix around its principal diagonal can obscure signal tendencies. To address this, we employ a modified recurrent plot technique for the temporal domain \cite{reference3}. This involves computing the angle between a reference vector and the temporal state difference vector \(s_m - s_n\), aiding in determining the sign of the recurrence plot in Equation \ref{eq:rp_matrix}:
\begin{equation}
\label{eq:temporal_matrix}
    R_{m,n} = sign(m,n) ||s_i - s_j||
\end{equation}

\subsection{Frequency Domain Enhancement in Recurrent Plots}
Building upon this, we integrate frequency domain data into the recurrent plot. We commence by postulating that state difference vectors in upward frequency phases predominantly lie in the first quadrant, while downward phases are in the third quadrant. After performing a Fourier transform on the temporal phases, we derive complex-valued frequency spectra, denoting each frequency component's phase as $p_i$ and $p_j$. We then calculate the angle between a base vector $v$ and the phase difference vector, utilizing a sign function for gradient direction differentiation. The sign function is defined as:
\begin{equation}
\text{sign}(m, n) = \begin{cases}
-1, & \text{if } \frac{(p^i-p^j).v}{||p^i-p^j||.||v||} <    \cos(\frac{3\pi}{4}) \\
1, & \text{otherwise}
\end{cases}
\end{equation}
where $v = [1, 1]$. This results in the frequency domain-adapted recurrence plot:
\begin{equation}
\label{eq:frequency_matrix}
    R_{m,n} = sign(m,n) ||p_i - p_j||
\end{equation}

For each target sensor channel (TEMP, EDA, and BVP), we apply Equation \ref{eq:frequency_matrix}, creating three distinct RP images. These images are then merged into a single matrix $M$ ($M \in \mathbb{R}^{(N-1) \times (N-1) \times 3}$), which is normalized and encoded as an RGB image.

\section{Experimental Evaluation}
\subsection{Dataset}
Our study leveraged a real-time data collected from 16 participants monitored over 8-10 days using invasive Dexcom G6 Continuous Glucose Monitors (CGM) and Empatica E4 wrist-worn devices \cite{reference2}. The Dexcom G6, an invasive needle incorporated glucose measuring device provided interstitial glucose readings every 5 minutes, while the Empatica E4 recorded multiple metrics including blood volume pulse (BVP), electrodermal activity (EDA), and skin temperature (TEMP) \cite{reference2}. As per the latest correlation study on blood glucose level with health vital features \cite{statistica1}, we selected TEMP, EDA, and BVP features from the wearable sensor readings they are sampled at 4Hz, 64Hz and 1Hz respectively.

\subsection{Preprocessing}
First, we combined the TEMP, BVP, and EDA sensor data into a three-dimensional structure, aligning them as x, y, and z dimensions respectively, and resampled the BVP data from 64Hz to match the 4Hz sampling rate of TEMP and EDA. We then extracted key information from the Dexcom CGP reading and synchronized these glucose readings with the 5-minute intervals of the biometric data. Finally, the biometric data was reshaped into a three-dimensional matrix with dimensions corresponding to the number of 5-minute samples, the number of readings per sample (1200, based on a 4Hz sampling rate), and the three biometric measurements, providing a comprehensive dataset for subsequent analysis.

\subsection{Encoding using Proposed RP}
In the next stage of our study, we concentrated on generating Recurrence Plots (RPs) from the structured data, utilizing these plots to visualize recurring patterns and analyze the dynamics within the time series data. For every 5-minute segment extracted from each participant's 3D matrix, we produced three separate RPs, each corresponding to one of the biometric indicators: TEMP, BVP, and EDA. Our approach encompassed phasic dimensions for the construction of these RPs. Subsequently, we normalized the RPs to standardize their matrix values within a predefined range. In the final step, we merged the individual RPs for TEMP, BVP, and EDA into a unified RGB image, assigning a distinct color channel in the RGB spectrum to each biometric reading. This process resulted in a collection of RGB images equivalent to the total count of 5-minute periods present in each participant's 3D matrix of TEMP, BVP and EDA readings, effectively capturing the composite biometric dynamics.

\subsection{Baselines Algorithms}
Our baseline for glucose prediction was established using methodologies from a study conducted by Duke University, which encompassed two distinct approaches: a gradient-boosted model for population-level analysis using leave-one-person-out cross-validation (LOPOCV), and a personalized gradient-boosted model tailored to individual data \cite{reference2}. In our study, we expanded the scope by incorporating two baseline methods: Lu et. al. proposed a modified temporal domain Recurrence Plot plot combined with a ResNet architecture (MTRP+ResNet) \cite{reference3}, and our proposed novel frequency domain Recurrence Plot plot also integrated with ResNet (MFRP+ResNet). 

    

\begin{table}[h]
\addtolength{\tabcolsep}{-4pt}
  \caption{Baseline algorithms' performance comparisons with our method }
  \label{tab:performance_comparison}
 \centering
 \begin{tabular}{|p{2.5cm}|p{0.9cm}|p{1.3cm}|p{1.6cm}|}
    \hline
    Method &  RMSE & MAPE & Accuracy(\%)\\
    \hline
    Population Model & 21.22 & 0.1433 & 85.67\\
    \hline
    Personalized Model &  21.10 & 0.1326 & 86.74\\
    \hline
    Our Method & {\bf 18.17} & {\bf 0.1235} & {\bf 87.65}\\
    \hline
  \end{tabular}
\end{table}

\subsection{Results Analysis}
In our study, we leveraged the capabilities of the ResNet-18 model, a renowned deep convolutional neural network with proven efficacy in image classification tasks. Our model's performance was rigorously evaluated using three distinct metrics: Mean Absolute Percentage Error (MAPE), Root Mean Squared Error (RMSE), and a bespoke accuracy metric calculated as (100 - MAPE)\%. To train the model, we adopted a composite loss function combining RMSE and MAPE. The training phase commenced with two varied datasets (Temporal and Phasic images) to probe the influence of different image representations on our predictive model's effectiveness. We partitioned the dataset into training and validation subsets in a 70:30 ratio. This division was crucial to maintain the integrity of the relationship between each Recurrence Plot image and its corresponding glucose level. Leveraging Dexcom datasets for individual participants, we meticulously mapped each image to its exact glucose measurement, ensuring precise data alignment.

The comparative analysis between our method and the Duke study is presented in Table 2. While both methodologies have their own merits, it's evident from the data that our method demonstrates superior performance in all three metrics evaluated. Specifically, our method achieved an RMSE of 18.17, which is a significant improvement over the Duke Study's 21.10. Additionally, our MAPE value was recorded at 0.1235, substantially better than the Duke Study's 0.1326. In terms of accuracy, our model, with an accuracy of 87.65\%, also marginally surpassed the Duke Study’s accuracy of 86.74\%. These improvements signify the effectiveness of our proposed model and its potential in real-world applications.

\subsection{Conclusion and Limitations}
In the face of rising global diabetes prevalence, the necessity for accurate and non-invasive glucose monitoring becomes crucial. Our research demonstrated the potential of utilizing wearables in conjunction with innovative data transformation techniques to surpass existing accuracy benchmarks in real-time glucose level prediction. 

Our exploration into the generation of Recurrence Plot (RP) Images in both temporal and phasic dimensions,has laid a solid foundation for further investigations in this domain. The process of converting raw biometric data into visual representations, specifically RGB images, has shown its merit in improving the data interpretability for deep learning models.

While our results are promising and outperform the benchmarks set by previous studies, it's essential to recognize the need for continued research. Future endeavors should focus on enhancing the model's robustness, expanding the dataset to include a more diverse demographic, and testing the model in real-world scenarios. Furthermore, an exploration into other potential biometrics and their integration could lead to even more accurate predictions.

\end{document}